\documentclass[lettersize,journal]{IEEEtran}

\usepackage{amsmath,amsfonts}
\usepackage{algorithmic}
\usepackage{algorithm}
\usepackage{array}
\usepackage[caption=false,font=footnotesize]{subfig}
\usepackage{textcomp}
\usepackage{stfloats}
\usepackage{url}
\usepackage{verbatim}
\usepackage{graphicx}
\usepackage{cite}
\usepackage[table]{xcolor}
 \usepackage{array}
 \usepackage{float}
\usepackage{graphicx}

\usepackage{array}
\usepackage{makecell}
\usepackage[hidelinks]{hyperref}
\begin{document}

\title{A Miniaturized Low-Cost 3-D Printed High-Permittivity Dielectric Transmissive LP-to-CP Converter for L-Band Satellite Communications\\
\thanks{This work was supported by the Deanship of Research at King Fahd University of Petroleum and Minerals (KFUPM) under project number EC251003. Corresponding author: Hussein Attia (e-mail: hattia@kfupm.edu.sa)}

}
\author{Gazali~Bashir,~\IEEEmembership{ Member,~IEEE}, 
        Javid Ahmad Ganie,~\IEEEmembership{ Member,~IEEE}, 
        Yazan Al-Alem,~\IEEEmembership{Senior Member,~IEEE}, 
        and Hussein Attia,~\IEEEmembership{Senior Member,~IEEE}%

        % <-this % stops a space
        \thanks{Gazali Bashir and Javid Ahmad Ganie are with the Interdisciplinary Research Center for
Communication Systems and Sensing, King Fahd University of Petroleum
\& Minerals (KFUPM), Dhahran 31261, Saudi Arabia (e-mail: gazali.bashir@kfupm.edu.sa)\\
Yazan Al-Alem and Hussein Attia are with the Department of Electrical Engineering and with the Interdisciplinary Research Center for
Communication Systems and Sensing, King Fahd University of Petroleum
\& Minerals (KFUPM), Dhahran 31261, Saudi Arabia.}%

}
\maketitle

\begin{abstract}
A low-cost, miniaturized, additively manufactured transmissive linear-to-circular polarization converter for L-band air and space communication systems is presented. The proposed polarizer employs an in-house-fabricated high-permittivity dielectric medium, realized using a 3-D printed polylactic acid (PLA) enclosure filled with a TiO$_2$--water mixture. Variation in the effective permittivity is achieved by adjusting the mixture filling fraction. The selected 60\% filling fraction corresponds to an effective relative permittivity of 11.3 at 1 GHz. The high-permittivity loading enables miniaturization of the polarizer height and area by reducing the anisotropic dielectric unit-cell dimensions to $\boldsymbol{0.037\lambda_0 \times 0.037\lambda_0 \times 0.22\lambda_0}$. An additively manufactured polarizer with overall dimensions $\boldsymbol{0.83\lambda_0 \times 0.77\lambda_0 \times 0.22\lambda_0}$ is experimentally validated using a compact linearly polarized feed source. The measured results demonstrate LP-to-CP conversion, with a 3 dB axial ratio of 12\% and insertion loss of less than 0.6 dB. Radiation-pattern measurements at 1 GHz show LHCP-to-RHCP discrimination levels of 19 and 24 dB in the \textbf{\textit{xz}}- and \textbf{\textit{yz}}-planes, respectively. The proposed configuration combines miniaturization, LP-to-CP conversion, and low-cost additive manufacturing, making it promising for UAV-mounted and other size-constrained L-band satellite communication terminals.
\end{abstract}


\section{Introduction}
Circularly polarized (CP) antennas are widely employed in satellite communication, global navigation satellite systems (GNSS), remote sensing, and synthetic-aperture radar owing to their ability to mitigate polarization-mismatch losses arising from relative antenna orientation and propagation-induced polarization rotation \cite{hester2024cubesat,ganie2025transmitarray,mahmoud2017wideband}.
 At L-band frequencies, relatively low atmospheric attenuation and reliable long-range propagation further support their use in satellite communication and navigation systems.

Circular polarization can be generated either directly by the radiating element or through an external polarization-conversion structure \cite{govindarajan2020compact,spaceAntennaHandbook}. Direct CP antennas generally excite two orthogonal modes with approximately equal amplitudes and a quadrature phase difference using dual-feed arrangements, perturbation elements, or specially shaped radiators \cite{gao2014circular}. Sequentially rotated elements and phase-feeding networks are also employed in CP antenna arrays \cite{hirasawa1992small}. However, such configurations increase the design complexity of the radiating structure and its excitation scheme. Alternatively, a linear-to-circular polarization (LP-to-CP) converter transforms the field radiated by a linearly polarized antenna into a CP wave without substantially modifying the feed antenna. This approach separates polarization control from radiation generation and enables existing linearly polarized antennas to be adapted for CP operation \cite{alalem2023millimeter}.

Transmissive LP-to-CP converters have been implemented using septum polarizers, meander-line structures, frequency-selective surfaces (FSSs), metasurfaces, and anisotropic dielectric structures \cite{wang2004septum,sofi2020frequency,yang2023linear}. Septum polarizers provide high polarization purity in waveguide systems but are not readily integrated with low-profile printed antennas. FSS- and metasurface-based converters offer planar configurations and considerable control over the transmitted amplitude and phase. Recent developments have substantially improved their bandwidth and angular stability. A multilayer K/Ka-band converter reported in \cite{geng2025broadband} achieves a measured 3-dB axial-ratio bandwidth of 52.8\% with an insertion loss below 1.65 dB. A single-substrate Huygen's metasurface exhibits a 64.2\% axial--ratio bandwidth from 16.3 to 31.7 GHz \cite{zhai2025high}. More recently, a dual-band K/Ka--band converter maintained LP-to-CP conversion over 19--21 and 29--31 GHz \cite{ouslimani2026kadual}. However, these structures employ patterned metallic resonators, multiple substrate layers, or impedance-transforming sections, increasing the fabrication and assembly complexity.

An alternative approach is to employ an anisotropic dielectric polarizer in which the two orthogonal components of an incident LP field experience different effective dielectric constants\cite{Hao_2020}. The resulting difference in propagation constants produces the required quadrature phase delay while avoiding metallic resonators within the polarizer. A dielectric polarizer reported in \cite{to2024xband} achieved a measured 3-dB axial-ratio bandwidth of 37.7\%, an insertion loss of approximately 0.5 dB, and a thickness of $0.21\lambda_0$ at X-band. Nevertheless, the design required a multilayer metamaterial configuration and additional impedance-matching layers.

Additive manufacturing has further extended the capabilities of dielectric structures by enabling lightweight and geometrically complex structures without requiring precision CNC milling, laser cutting, or high-temperature ceramic processing\cite{javid}. A 24--GHz antenna array integrated with a 3--D printed dielectric polarizer demonstrates a gain of 20 dBic and a 3-dB axial-ratio bandwidth of 6\% \cite{alalem2023millimeter}. An air–dielectric-slab polarizer integrated with a 60 GHz radiating aperture achieves a 30\% overlapping impedance–axial-ratio bandwidth, and a peak gain of 15 dBic \cite{wang2017wideband}. A low-cost millimeter-wave antenna with a peak gain of  15 dBic, a 30\% axial-ratio bandwidth, and a 20\% overlapping impedance-axial-ratio bandwidth is presented in \cite{alalem2022lowcost}. A Ka-band antenna combining a printed ridge gap waveguide fed with a 3-D-printed dielectric polarizer achieved a gain of 20 dBi and a 3-dB axial-ratio bandwidth of 6.5\% \cite{alalem2023prgw}. A spherical dielectric polarization converter with an axial-ratio bandwidth of  34\%  is presented in \cite{zhao2021conformal}. Logarithmic-spiral and spherically conformed dielectric polarizers have also been investigated for dual-CP conical and omnidirectional radiation in the Ka-band \cite{melendro2024logarithmic,melendro2025loxodromic}.

Despite these developments, most additively manufactured dielectric polarizers operate at X, Ku, Ka, or higher millimeter-wave frequencies and employ commercially available materials with fixed and relatively low permittivities. Directly scaling such structures to the L-band results in larger physical dimensions due to the longer wavelength. Although high-permittivity dielectrics can reduce the required dimensions, commercially available high-permittivity materials are often expensive, difficult to shape, or provide limited flexibility for dielectric-property customization. To address this requirement, this paper presents a miniaturized transmissive LP-to-CP converter based on a high-permittivity dielectric medium. The medium consists of a 3-D printed PLA enclosure filled with a TiO$_2$–water mixture and exhibits an effective relative permittivity of approximately 11.3 near 1 GHz.\\
The main contributions of this work are summarized as follows:
\begin{enumerate}
    \item Low-cost in-house fabricated high-permittivity medium: The dielectric medium comprises a 3-D-printed PLA enclosure filled with a TiO$_2$--water mixture and exhibits an effective relative permittivity of 11.3 at 1 GHz without commercial high-permittivity substrates or specialized fabrication processes.

    \item Effective Permittivity Adjustment: The effective relative permittivity is varied by adjusting the TiO$_2$–water mixture filling ratio within the PLA, allowing the dielectric response to be tailored to the polarizer design requirements while retaining the same PLA enclosure.
    
    \item Low-profile all-dielectric configuration: The anisotropic unit cell generates the required quadrature phase delay through dielectric without loading  patterned metallic
resonators. The polarizer has a thickness of \(0.22\lambda_0\), with overall dimensions of \(0.83\lambda_0 \times 0.77\lambda_0 \times 0.22\lambda_0\).

    \item Wideband polarization conversion and high polarization discrimination: The polarizer exhibits a 12\% 3-dB axial-ratio bandwidth and provides LHCP-to-RHCP discrimination levels of 19 and 24 dB in the $xz$- and $yz$-planes, respectively.
\end{enumerate}

\section{ High Relative Permittivity Material Design}
\begin{figure}[t]
\centering
{%
\includegraphics[width=0.45\textwidth]{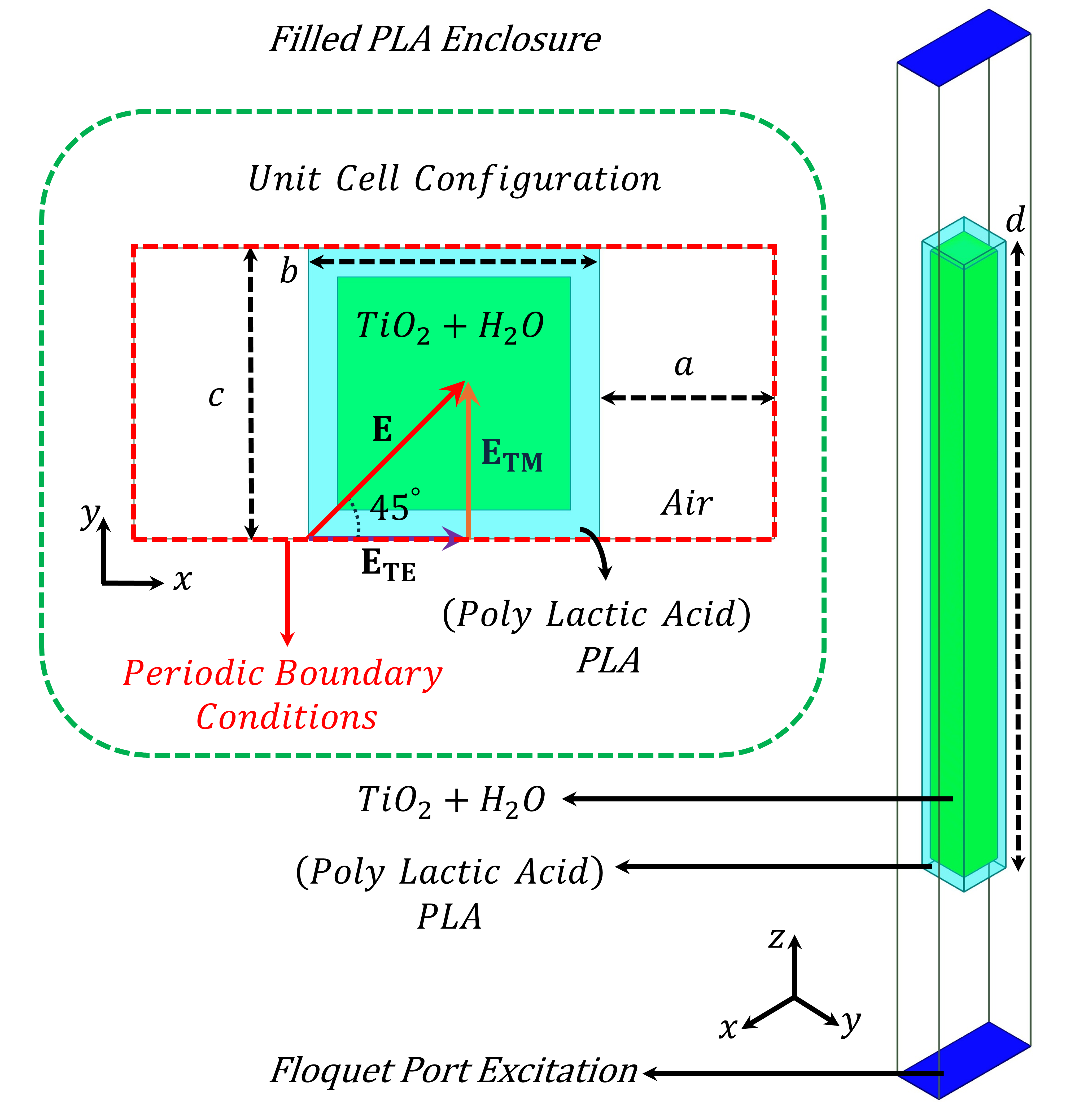}%
}
\vspace{-.25cm}
\caption{Proposed permittivity-controlled dielectric unit cell.}
\label{fig01}
\vspace{-.3 cm}
\end{figure}

\begin{figure}[!h]
\centering
\subfloat[]{%
\includegraphics[width=0.24\textwidth]{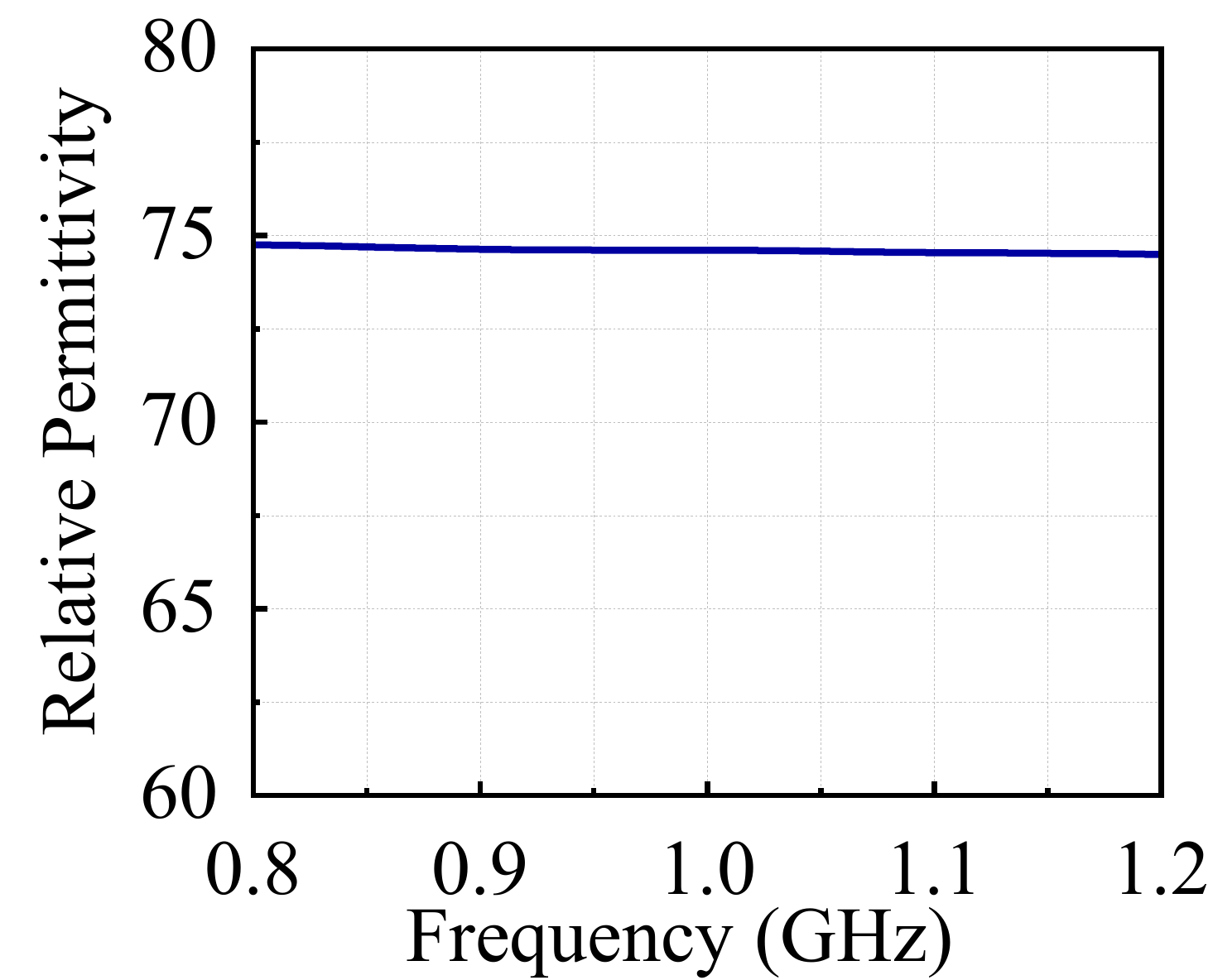}%
}
\hfill
\subfloat[]{%
    \includegraphics[width=0.24\textwidth]{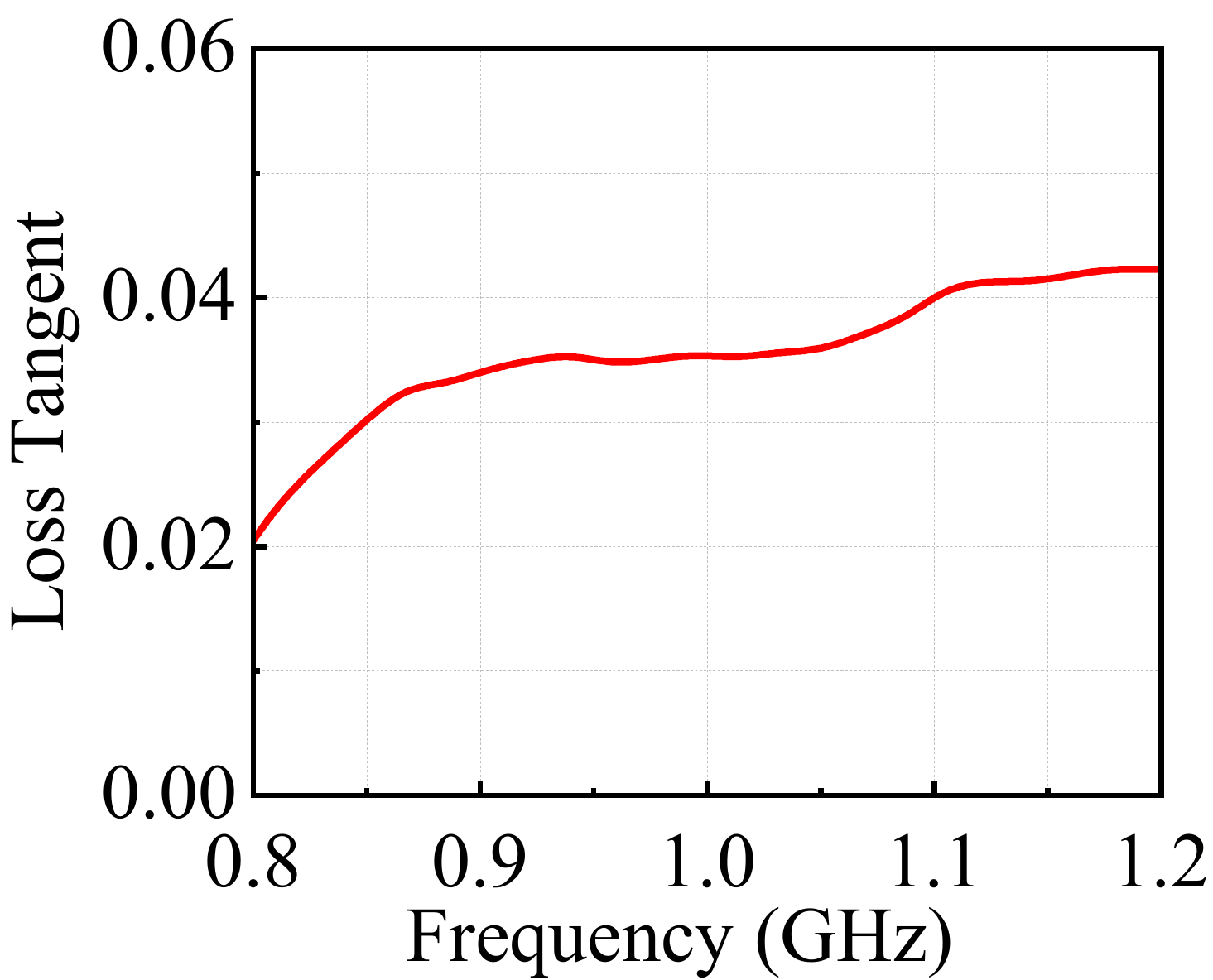}%
}
\caption{Dielectric characterization of the TiO$_2$–water mixture. (a) measured relative permittivity, and (b) loss tangent.}
\label{fig02}
\vspace{-.4cm}
\end{figure}

To obtain the high permittivity required for L-band miniaturization, a TiO$_2$–water mixture is incorporated into a 3D-printed PLA enclosure, as illustrated in Fig.~\ref{fig01}. The mixture contains equal proportions of TiO$_2$ and water. 
%%%%%%%%%%%%%%%
%%%%%%%%%%%%%%%
Its dielectric properties are characterized using a Keysight N1501A dielectric probe kit over the frequency range 0.8–1.2 GHz. Fig.~\ref{fig02}(a) and Fig.~\ref{fig02}(b) present the measured relative permittivity and loss tangent. At 1 GHz, the mixture exhibits a relative permittivity of 74.5 and a loss tangent of 0.035. The relative permittivity decreases with frequency, while the loss tangent increases.

\begin{figure}[b]
\centering
{%
\includegraphics[width=0.44\textwidth]{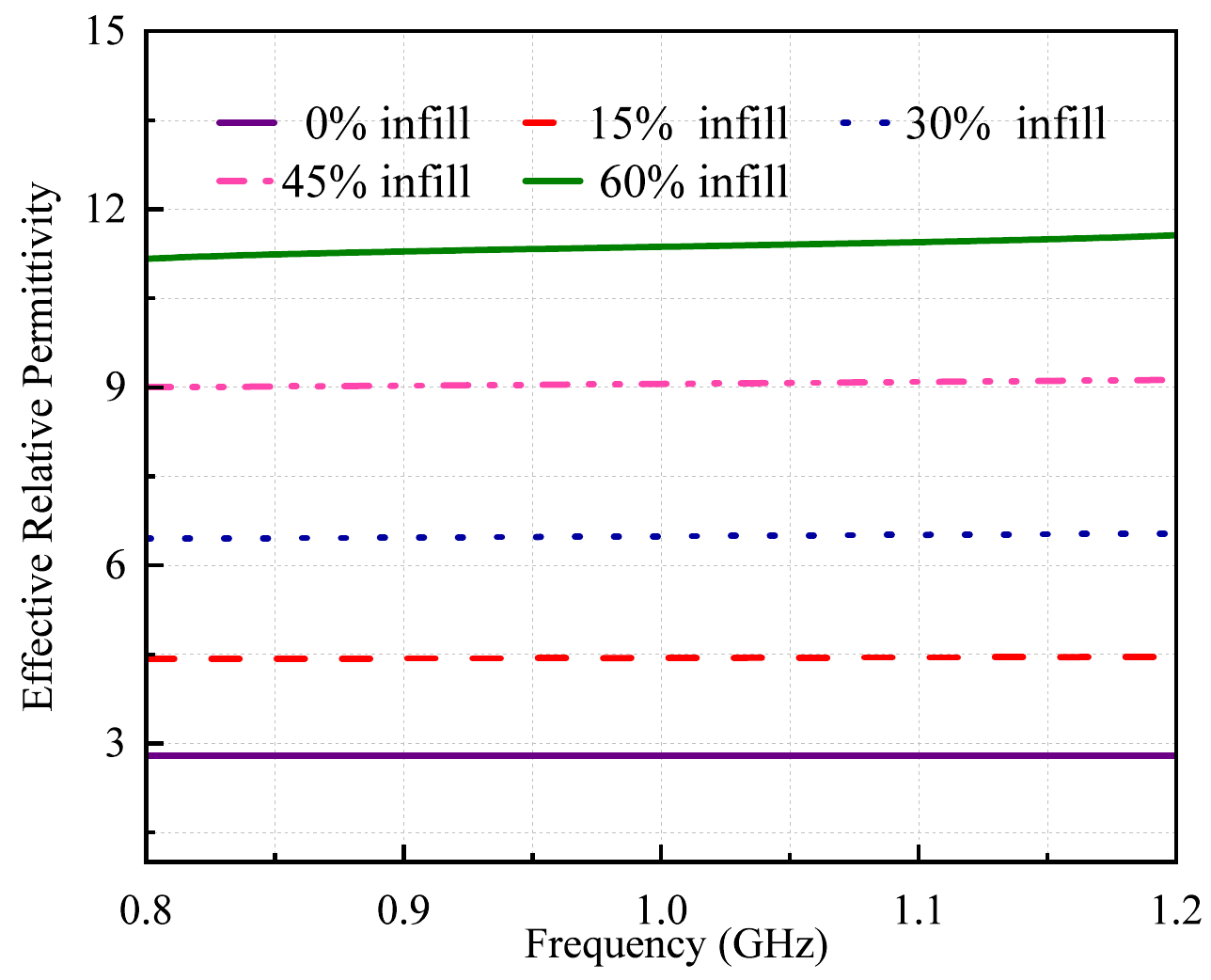}%
}
\vspace{-.22cm}
\caption{Extracted effective relative permittivity for different mixture filling ratios.}
\label{fig022}
\end{figure}

%%%%%%%%%%%%%%%%%%
The measured frequency-dependent dielectric properties are assigned to the filling region of the PLA enclosure. The effective relative permittivity of the combined PLA/TiO$_2$--water unit cell is then extracted in CST Microwave Studio, where the PLA is modeled with $\varepsilon_r = 2.8$ and the unit cell is simulated under periodic boundary conditions with Floquet ports. The effective permittivity is retrieved from the simulated S-parameters. Varying the TiO$_2$--water mixture filling fraction changes the effective dielectric response of the combined PLA/TiO$_2$--water unit cell. Fig.~\ref{fig022} presents the extracted effective relative permittivity for filling fractions ranging from 0\% to 60\%. The dependence of the effective relative permittivity on the filling fraction enables permittivity adjustment without modifying the PLA enclosure geometry. The selected filling fraction of 60\% yields an effective relative permittivity of 11.3 at 1~GHz and maintains a stable response across the operating band. Although the effective permittivity of the composite is not measured directly, it is validated indirectly through the measured quadrature phase difference and axial ratio of the fabricated polarizer in Section~V. This configuration is subsequently employed in the LP-to-CP unit-cell design.
% \vspace{-.5cm}
\section{Dielectric Polarizer Unit-Cell Design}
The proposed LP-to-CP converter is realized using a transmissive anisotropic dielectric unit cell periodically arranged to form a finite dielectric polarizer. The polarization conversion mechanism relies on the introduction of different effective dielectric constants for the two orthogonal field components propagating through the dielectric structure. Consequently, the orthogonal components experience different propagation velocities, resulting in the required phase delay for linear-to-circular polarization conversion while maintaining high transmission efficiency. Fig.~\ref{fig01} illustrates the proposed dielectric unit cell, comprising a centrally positioned composite dielectric column and two lateral air regions.
The composite column consists of a PLA enclosure with a relative permittivity of 2.8 and a TiO$_2$--water mixture occupying 60\% of the dielectric column volume. The measured frequency-dependent dielectric properties are
assigned to the TiO$_2$--water mixture. An effective relative
permittivity of 11.3 is obtained at 1 GHz for the selected
configuration, with a stable response over the polarizer
operating band, as shown in Fig.~\ref{fig022}. The dielectric-column width is denoted by \(b\), the width of each lateral air region by \(a\), and the unit-cell period along the \(x\)-direction is \(2a+b\). The propagation
length along the $z$-direction is denoted by $d$.
For an incident linearly polarized wave oriented at $45^\circ$, the electric field can be decomposed into two orthogonal components, transverse-electric (TE) and transverse-magnetic (TM) components. Due to the anisotropic dielectric distribution, the two components experience different effective dielectric constants while propagating through the unit cell. Following the effective-medium approximation, the equivalent
dielectric constants can be expressed by \eqref{eq:epsx} and
\eqref{eq:epsy} \cite{Wang2015Wideband}.
\begin{equation}
\varepsilon_{\mathrm{eff},x} \approx 1
\label{eq:epsx}
\end{equation}
\begin{equation}
\varepsilon_{\mathrm{eff},y} \approx 1 + (\varepsilon_r-1)\frac{{b}}{2a + b}
\label{eq:epsy}
\end{equation}
where $\varepsilon_r$ denotes the relative permittivity of the proposed dielectric medium. The resulting phase difference between the two orthogonal components can be approximated using \eqref{eq:phase}.
%%%%%%%%%%%%%%%%%%%%%%%%%%%%%%
\begin{figure}
\centering
\subfloat[]{\includegraphics[width=0.24\textwidth]{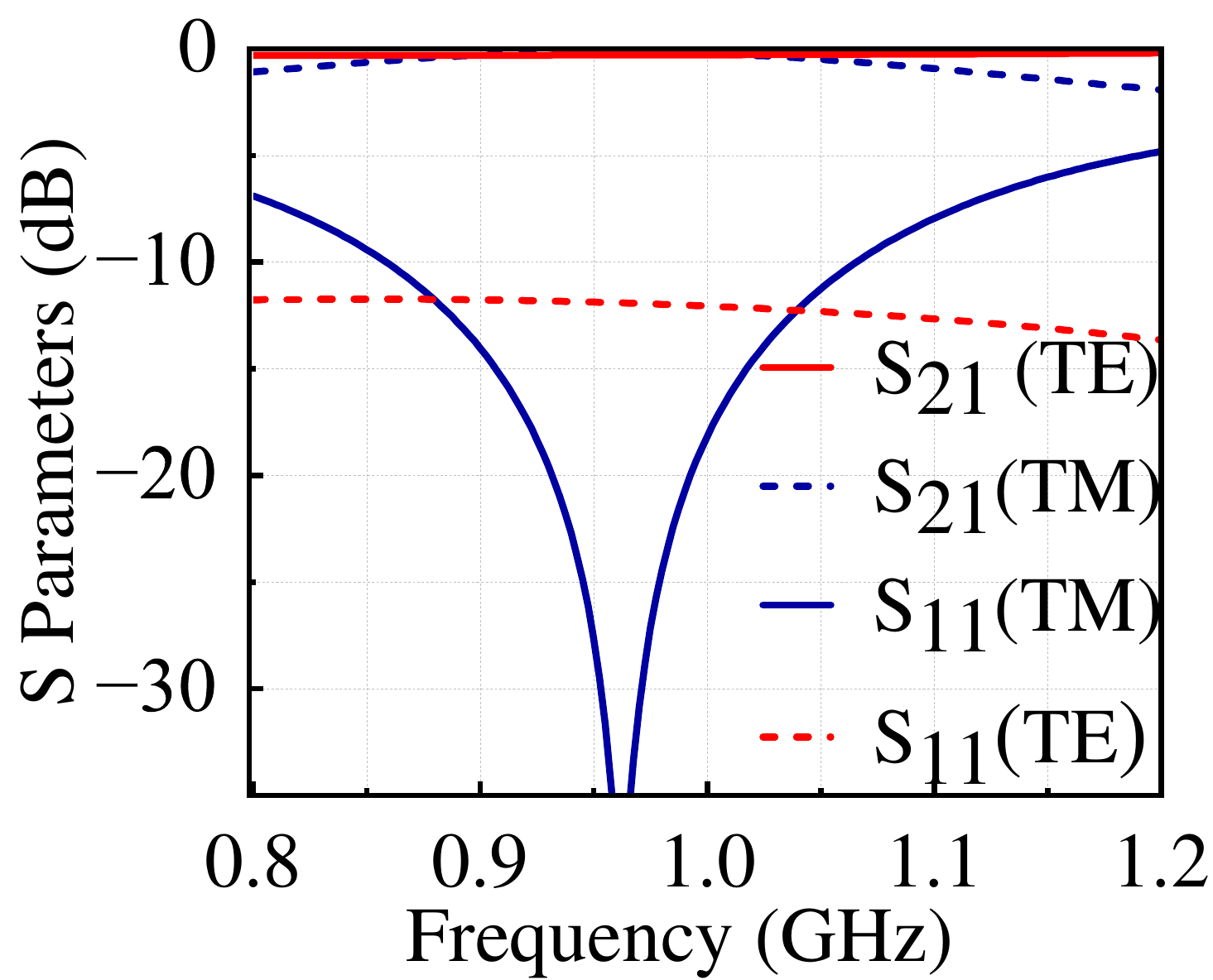}}
\label{3a}
\hfill
\subfloat[]{\includegraphics[width=0.24\textwidth]{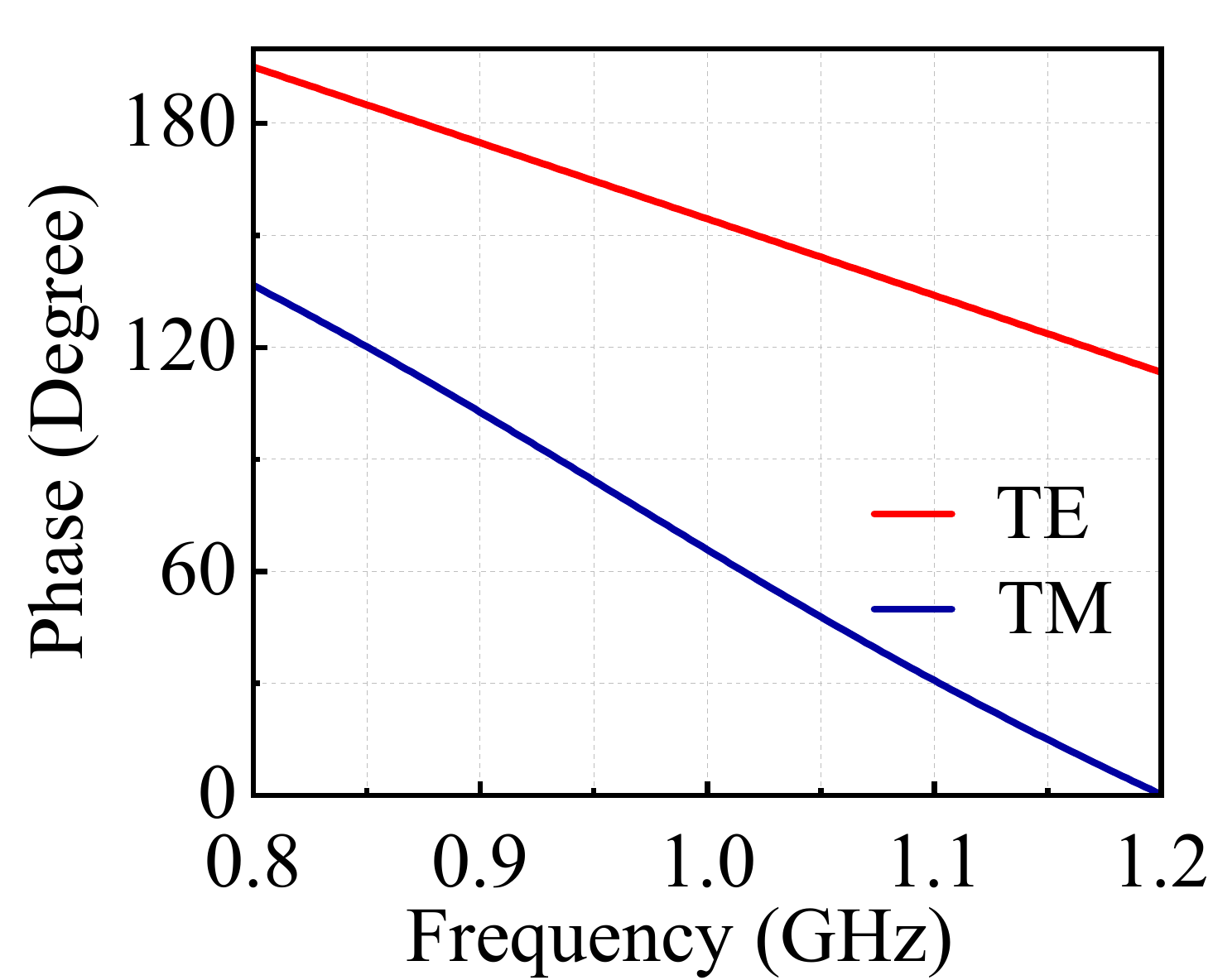}}%
\label{3b}
 \hfill 
 \subfloat[]
{\includegraphics[width=0.24\textwidth]{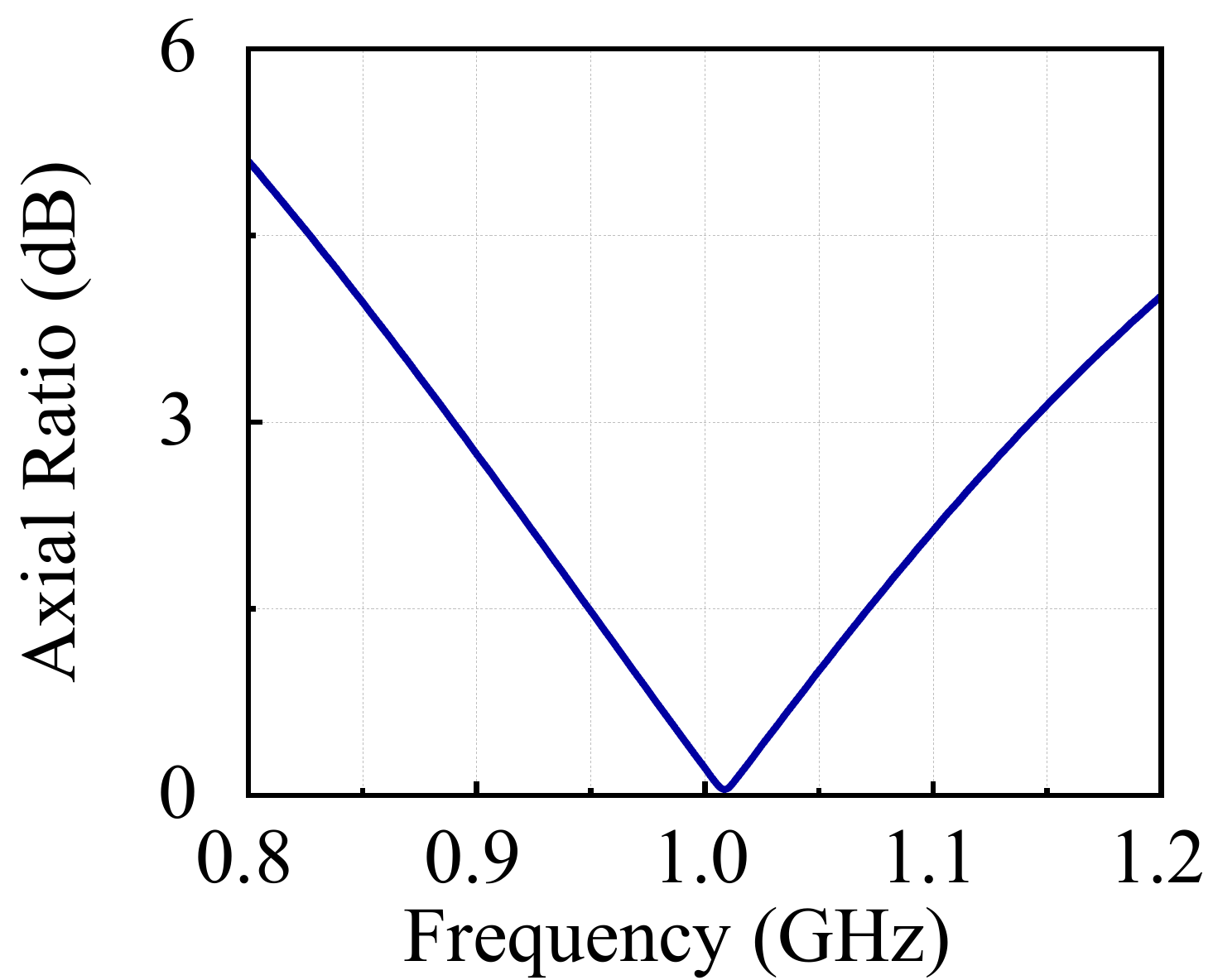}}
\label{3c}
\hfill
 \subfloat[]
{\includegraphics[width=0.24\textwidth]{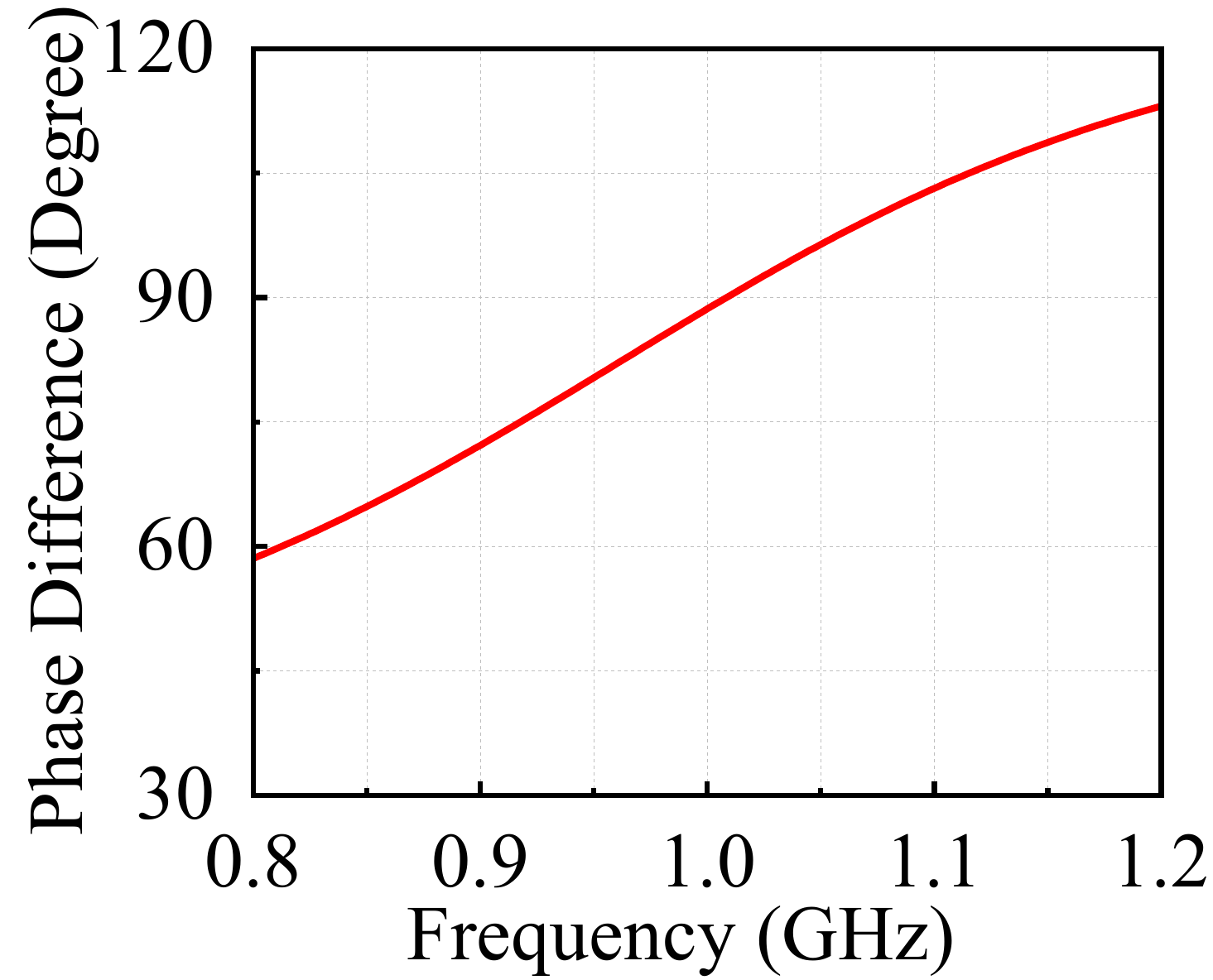}}
\label{dilec}
\caption{Simulated unit-cell response. (a) transmission and reflection coefficients, (b) transmission phase, (c) axial ratio, and (d) phase difference between TE and TM modes.}
\label{fig21}
\end{figure}

\begin{equation}
\Delta\theta=
\frac{2\pi f}{c_{o}}
\left(
\sqrt{\varepsilon_{\mathrm{eff},y}}-1
\right)d
\label{eq:phase}
\end{equation}
where $f$ denotes the operating frequency, $c_0$ is the speed of light
in free space, and $d$ is the propagation length of the dielectric
unit cell. The corresponding axial
ratio can be estimated using \eqref{eq:AR}.
\begin{equation}
\mathrm{AR}_{\mathrm{dB}}=
\left|
20\log_{10}
\left[
\tan\left(\frac{\Delta\theta}{2}\right)
\right]
\right|
\label{eq:AR}
\end{equation} Circular polarization is achieved when the transmitted orthogonal components exhibit approximately equal amplitudes together with a phase difference close to $90^\circ$.
\begin{figure}[!t]
\centering

\subfloat[]{%
    \includegraphics[width=0.97\columnwidth]{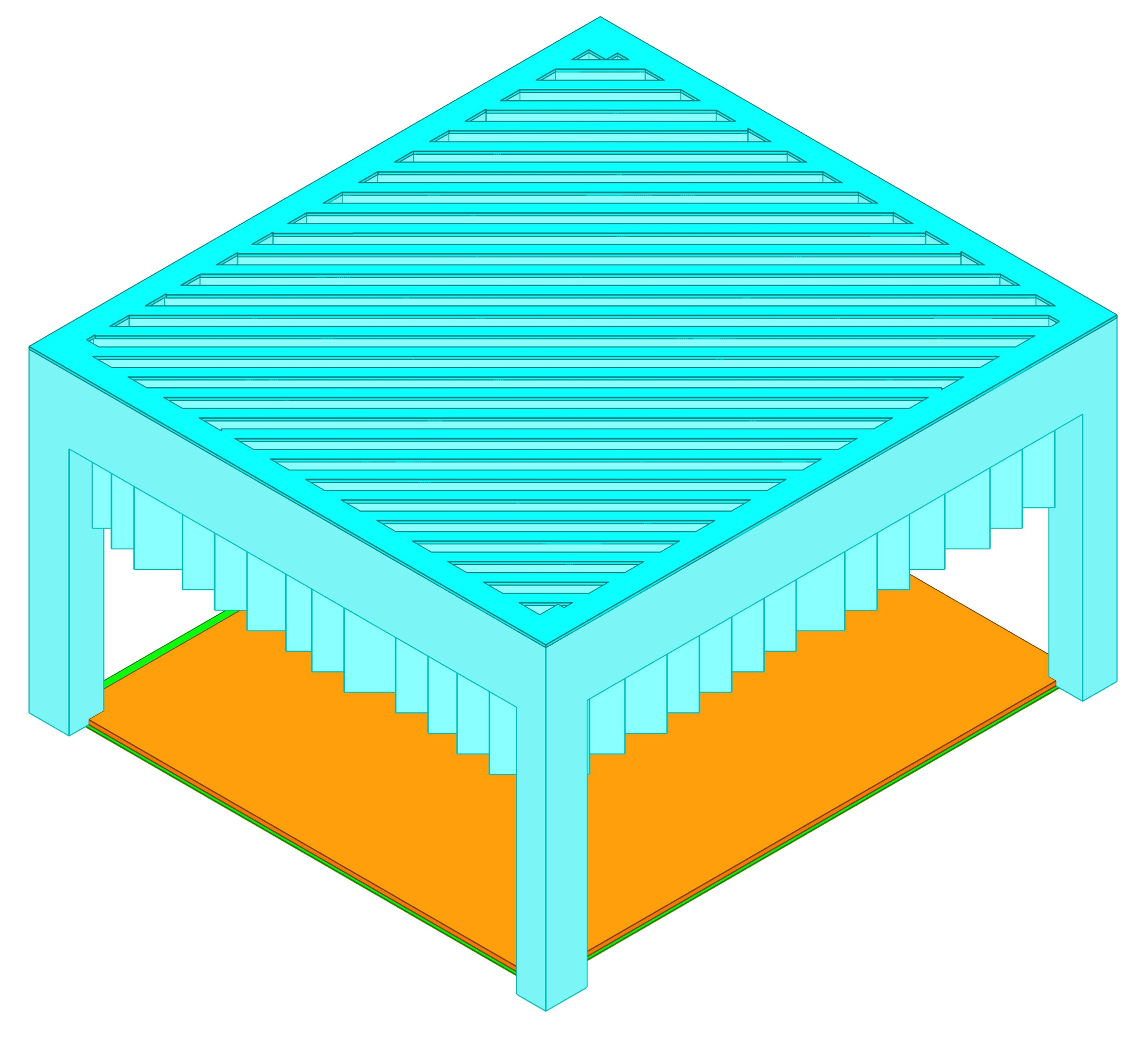}%
}

\par\vspace{1mm}

\centering
\subfloat[]{%
    \includegraphics[width=0.967\columnwidth]{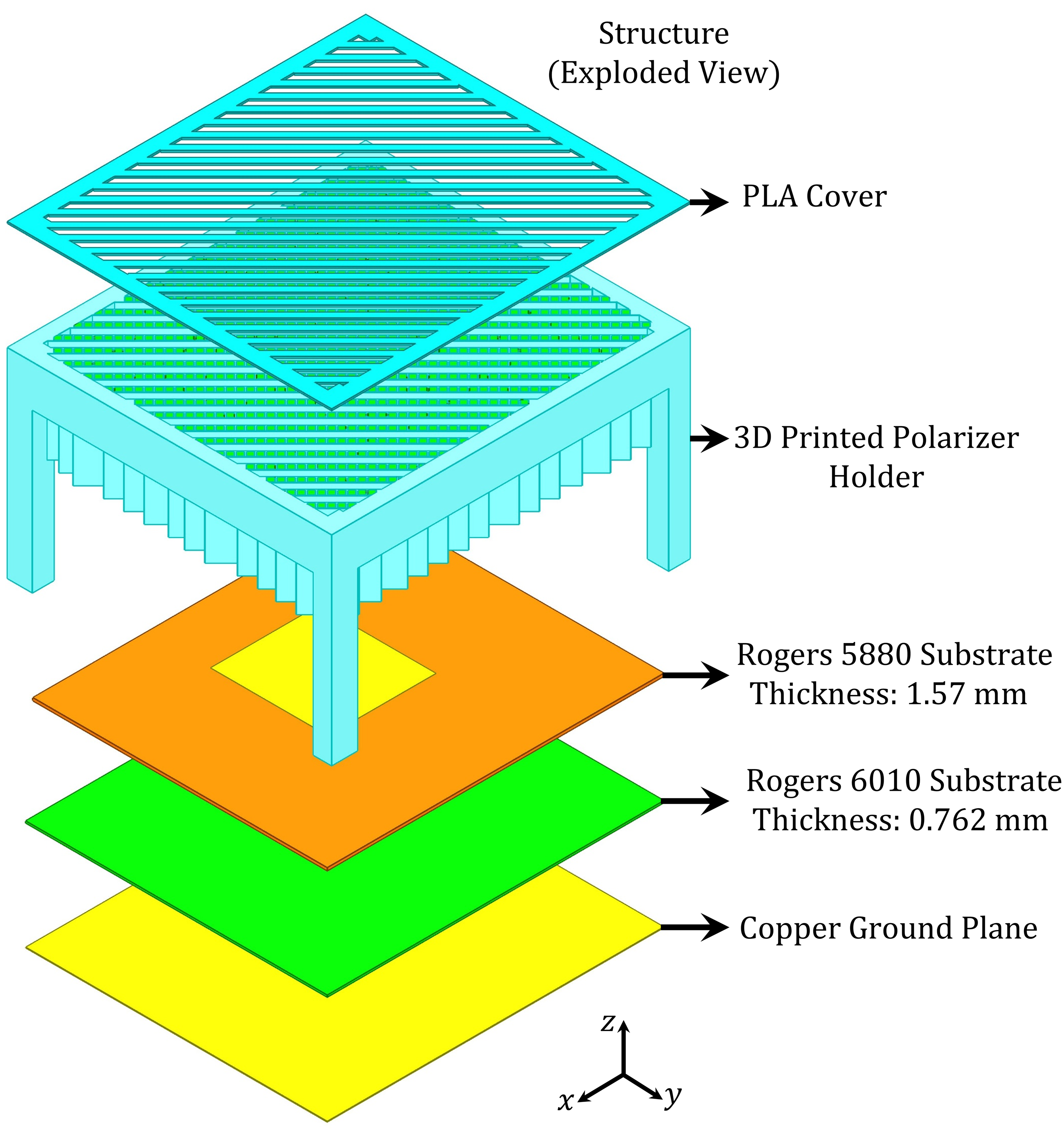}%
}
\caption{Configuration of the proposed linear to circular polarization converter (a) assembled feed--polarizer structure and (b) exploded view.}
\label{fig:f045}
\vspace{-.2cm}
\end{figure}
%%%%%%%%%%%%%%%%
The unit cell is analyzed using the frequency-domain solver in CST Microwave Studio, with unit-cell boundary conditions applied along the $x$- and $y$-directions, and Floquet ports assigned along the $z$-direction. Fig.~\ref{fig21}(a) shows the simulated transmission and reflection coefficients of the optimized unit cell under periodic boundary conditions. It can be observed that both TE and TM components exhibit transmission coefficients close to 0 dB over the operating band, indicating low insertion loss and high transmission efficiency. Furthermore, the reflection coefficients remain below approximately -10 dB across the frequency range of interest, confirming good impedance matching and minimal reflected power. Fig.~\ref{fig21}(b) shows the transmission phases of the TE and TM components. The phase difference between the TE and TM components is shown in Fig. 4(d), indicating that the required phase delay approaches \(90^\circ\) within the operating band. The simulated axial-ratio response of the optimized unit cell is shown in  Fig.~\ref{fig21}(c). The proposed unit cell achieves a simulated 3-dB axial-ratio bandwidth from 0.89 to 1.14 GHz. 
\begin{figure}[t]
\centering
{%
\includegraphics[width=0.92\columnwidth]{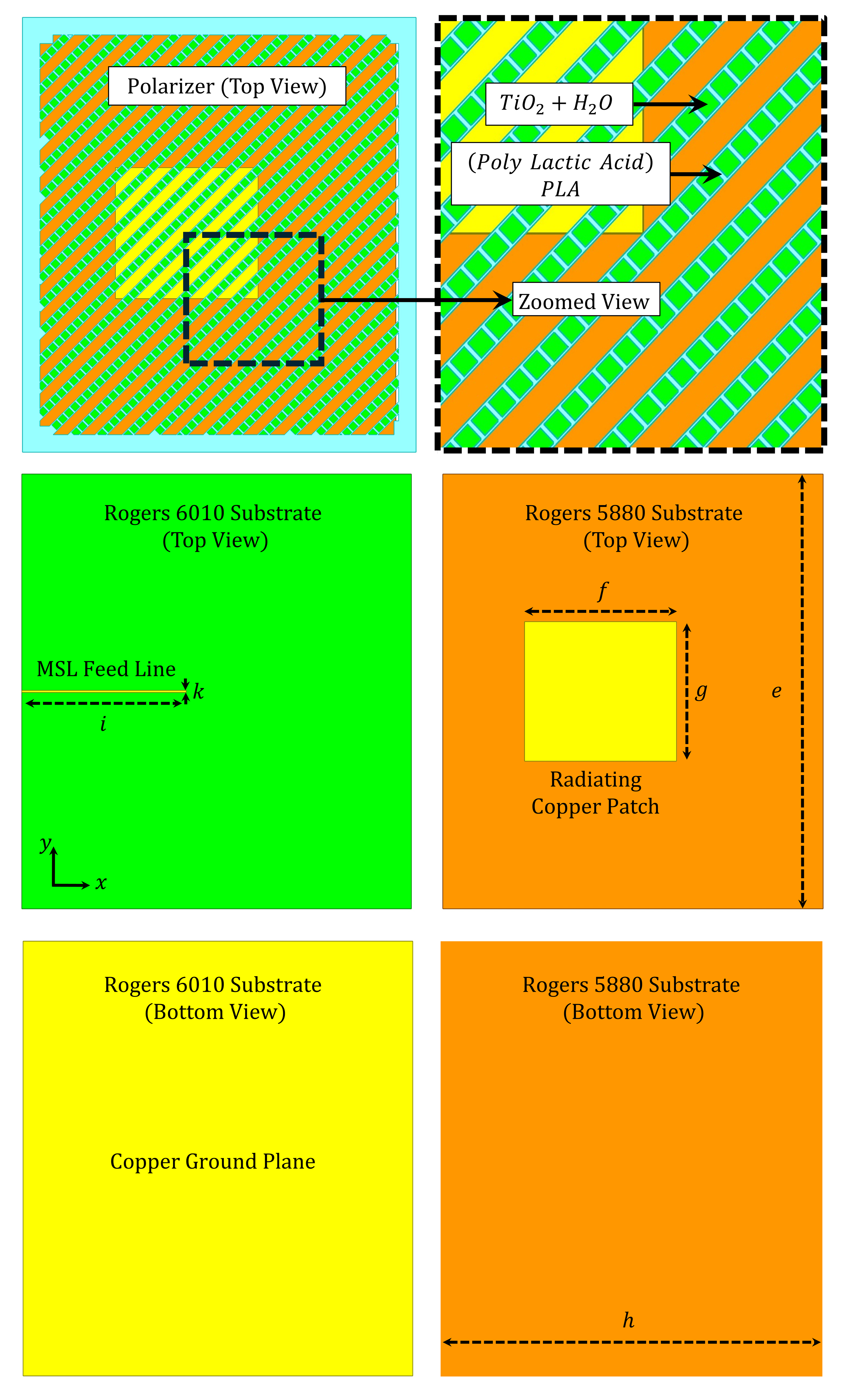}%
}
\caption{Layer configuration of the proposed polarizer and feed source.}
\label{fig0262}
% \vspace{-.45cm}
\end{figure}
\section{Linear to Circular Polarization  Converter Configuration}
To validate the LP-to-CP conversion of the converter, a compact linearly polarized feed source is employed as the excitation source, as shown in  Fig.~\ref{fig:f045} and Fig.~\ref{fig0262}. The feeding source consists of two dielectric substrate layers. The bottom substrate, which supports the ground plane and the microstrip feed line, printed on  Rogers  6010 LM with a relative permittivity of $\varepsilon_r=10.2$, a loss tangent of 0.0023, and a thickness of 0.762~mm. The radiating feed is printed on the upper substrate, which is Rogers RT/Duroid 5880 with a relative permittivity of $\varepsilon_r=2.2$, a loss tangent of 0.0009, and a thickness of 1.57~mm. Electromagnetic coupling between the feed line and the radiating feed is controlled by the overlap region, thereby enabling excitation of the radiating element.
The dielectric polarizer is realized by arranging the optimized dielectric unit cells into a finite surface with overall dimensions of $0.83\lambda_0 \times 0.77\lambda_0 \times 0.22\lambda_0$. To achieve circular polarization, the dielectric polarizer is rotated $45^\circ$ relative to the feed source linear polarization, ensuring equal excitation of the TE and TM components. The  dimensions of the polarizer
 and the feed source are listed in Table~I.
 \begin{figure}[!t]
 \vspace{-.3cm}
\centering
{%
    \includegraphics[width=0.92\columnwidth]{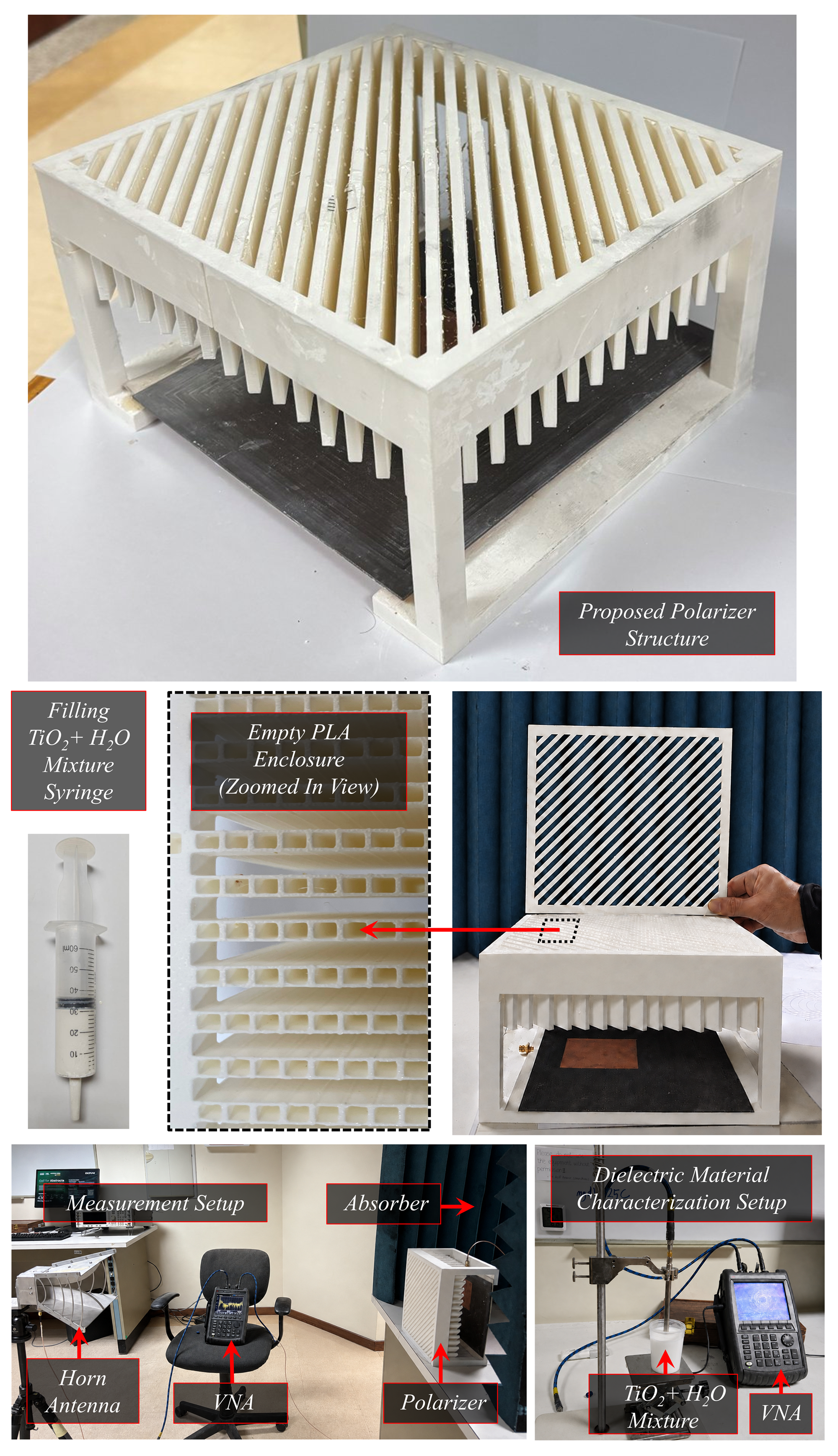}%
}
\caption{Fabricated prototype and the measurement setup.}
\label{f05}
\end{figure}
\begin{table}[t]
\caption{Dimensions of the Polarizer Structure and Feed in (MM)}
\label{tab:polarizer_dimensions}
\centering
\footnotesize
\renewcommand{\arraystretch}{0.95}

\begin{tabular*}{0.9\columnwidth}{
@{\extracolsep{\fill}}*{11}{c}@{}}
\hline
\textbf{\textit{a}} &
\textbf{\textit{b}} &
\textbf{\textit{c}} &
\textbf{\textit{d}} &
\textbf{\textit{e}} &
\textbf{\textit{f}} &
\textbf{\textit{g}} &
\textbf{\textit{h}} &
\textbf{\textit{i}} &
\textbf{\textit{j}} &
\textbf{\textit{k}} \\
\hline
3 & 5 & 11 & 65 & 230 & 75 & 83 & 250 & 102.70 & 80 & 1.4 \\
\hline
\end{tabular*}
 \vspace{-.1cm}
\end{table}
%%%
%%%%%%%%
\section{Prototype Fabrication and Experimental Validation}
A prototype of the proposed dielectric polarizer is fabricated
to experimentally validate the LP-to-CP polarization conversion
performance. The PLA enclosure elements and supporting-frame components
are fabricated by fused deposition modeling (FDM) using a Bambu Lab X1E
3-D printer. The printed components are subsequently assembled to form
a $250 \times 230 \times 65~\mathrm{mm}^{3}$ dielectric polarizer. The additive-manufacturing process enables the anisotropic enclosure geometry to be realized without metallic patterning or precision machining and allows the dielectric response to be adjusted through the mixture filling fraction while retaining the same printed geometry.
\begin{figure}[t]
\centering
\includegraphics[width=\columnwidth]{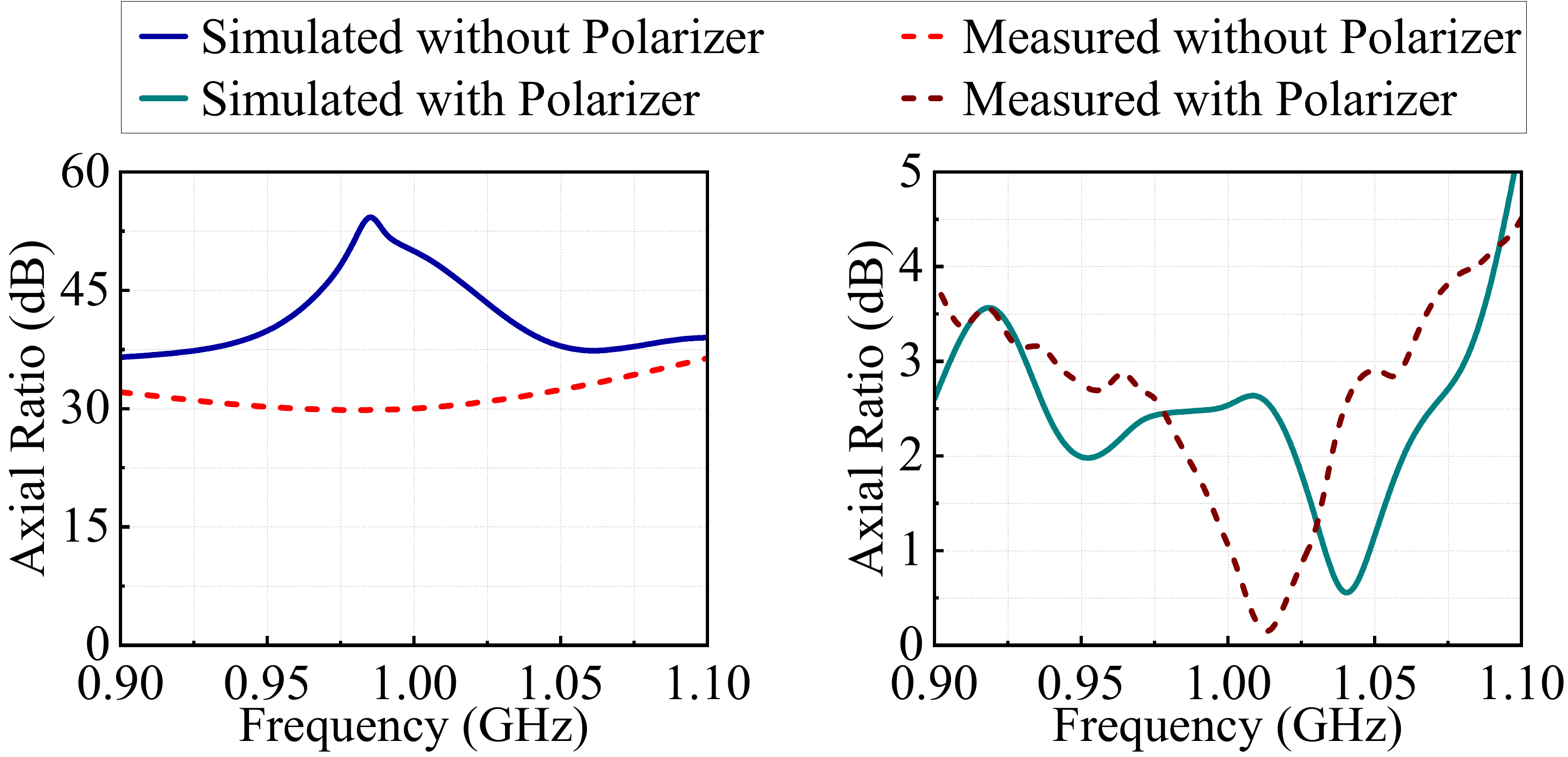}
\par\vspace{-1mm}

\makebox[\columnwidth][c]{%
    \makebox[0.5\columnwidth][c]{\footnotesize (a)}%
    \makebox[0.5\columnwidth][c]{\footnotesize (b)}%
}
\par\vspace{-1mm}

\caption{Simulated and measured axial-ratio of (a) the feed source and (b) the feed source with the dielectric polarizer.}
\label{axrl}
\end{figure}
Following assembly, the internal cavities are filled with the TiO$_2$--water mixture at the selected filling fraction of 60\% and sealed to minimize leakage and water evaporation. Since water dominates the permittivity of the TiO$_2$--water mixture, any evaporation or temperature variation would alter the effective permittivity of the dielectric column and shift the frequency at which the quadrature phase delay in (3) is achieved. For long-term deployment in harsh environments, a solid or gelled high-permittivity filler, such as a TiO$_2$--resin composite, can be adopted, which is the subject of future work. The feed source is fabricated using standard printed circuit board technology. The lower Rogers RT/duroid 6010.2LM substrate supports the ground plane and microstrip feed line, while the radiating patch is printed on the upper Rogers RT/duroid 5880 substrate. The supporting frame positions the polarizer 60~mm above the radiating patch, with its principal axis oriented $45^{\circ}$ relative to the linear polarization axis of the feeding source. The assembled prototype is maintained at room temperature until thermal stabilization is achieved, and the measurements are subsequently conducted under controlled laboratory conditions. The fabricated polarizer, its integration with the feed, and the measurement setup are shown in Fig.~\ref{f05}. The axial ratio and radiation patterns are characterized in the far field using a Keysight N9918A vector network analyzer. The simulated and measured axial-ratio responses of the standalone feed and the feed integrated with the polarizer are compared in Fig.~\ref{axrl}. The standalone feed exhibits an axial ratio above 25~dB, confirming its linear polarization, whereas the feed integrated with the polarizer achieves a measured 3-dB axial-ratio bandwidth of 12\%, from 0.94 to 1.06~GHz. The difference between the simulated and measured axial-ratio bandwidths is mainly attributed to fabrication tolerances of the FDM-printed enclosure, slight variations in the filling level and homogeneity of the TiO$_2$--water mixture, and alignment tolerances in the feed--polarizer spacing and the $45^{\circ}$ orientation. \\The radiation patterns of the feed source in Fig.~\ref{ffpp} exhibit broadside radiation, with sidelobe and cross-polarization levels below $-12$~dB in both principal planes, thereby providing suitable illumination of the polarizer along its surface normal for LP-to-CP conversion.
%%%%%%%%%%%%%%%%%
Fig.~\ref{fig9} shows the simulated and measured normalized LHCP and
RHCP radiation patterns of the polarizer at 1~GHz. Broadside
LHCP radiation is observed in both principal planes, with measured
sidelobe levels below $-14$ and $-15$~dB in the $xz$- and
$yz$-planes, respectively. At broadside, the measured RHCP levels are
19 and 24~dB below the corresponding LHCP levels in the $xz$- and
$yz$-planes, respectively, confirming the LP-to-CP polarization
conversion. Fig.~\ref{lss} presents the measured transmission characteristics with and without the polarizer at an antenna separation of 50 cm. With the polarizer, the (Horizontal) H- and (Vertical) V-polarized components exhibit nearly equal magnitudes and an approximately \(90^\circ\) phase difference over the operating band, confirming LP-to-CP conversion. The additional insertion loss introduced by the polarizer remains below 0.6 dB, demonstrating its low-loss transmission performance. The simulated and measured reflection coefficients of the
standalone feed  and the polarizer are compared in
Fig.~\ref{direc}(a). The
 simulated and measured directivity responses are presented
in Fig.~\ref{direc}(b).

The performance of the proposed polarizer is compared with other transmissive LP-to-CP converters, as summarized in Table~II. To compare designs operating at different frequencies, the polarizer height is normalized to $\lambda_0$ at the center frequency. This is particularly important at L-band, where the long wavelength translates even moderate normalized heights into large physical dimensions. The proposed polarizer has the lowest profile ($0.22\lambda_0$, 65~mm) among those reported in \cite{geng2025broadband,ouslimani2026kadual,zhao2021conformal,ding2021lowprofile, Dicandia2022, Arnieri2021}, while maintaining the lowest insertion loss (0.6~dB) among the compared works. Unlike the multilayer or metallic configurations in \cite{geng2025broadband,ouslimani2026kadual,to2024xband,yang2024ultrawideband}, the proposed polarizer employs a single additively manufactured all-dielectric surface without metallic patterning or additional impedance-matching layers, thereby reducing fabrication and assembly complexity, which makes it well suited to size-constrained L-band terminals such as UAV-mounted systems                       .
%%%%%%%%%%%%%%%%%%%
% \begin{figure}
% \centering
% \subfloat[]{\includegraphics[width=0.241\textwidth]{imagef/ex.pdf}}
% \hfill
% \subfloat[]{\includegraphics[width=0.247\textwidth]{imagef/hx.pdf}}
% \caption{ Simulated and measured axial-ratio responses. (a) probe, and (b) probe with a dielectric polarizer.}
% \label{xpp}
% \end{figure}
%%%%%%%%%%%%%%%%%
%%%%%%%%%%%%%%%%%%%%%%%%%%%%
%%

%%%%%%%%%%%%%%%%%%%%%%%
%%%%%%%%%%%%%%%%
\begin{figure}[t]
\centering
\includegraphics[width=\columnwidth]{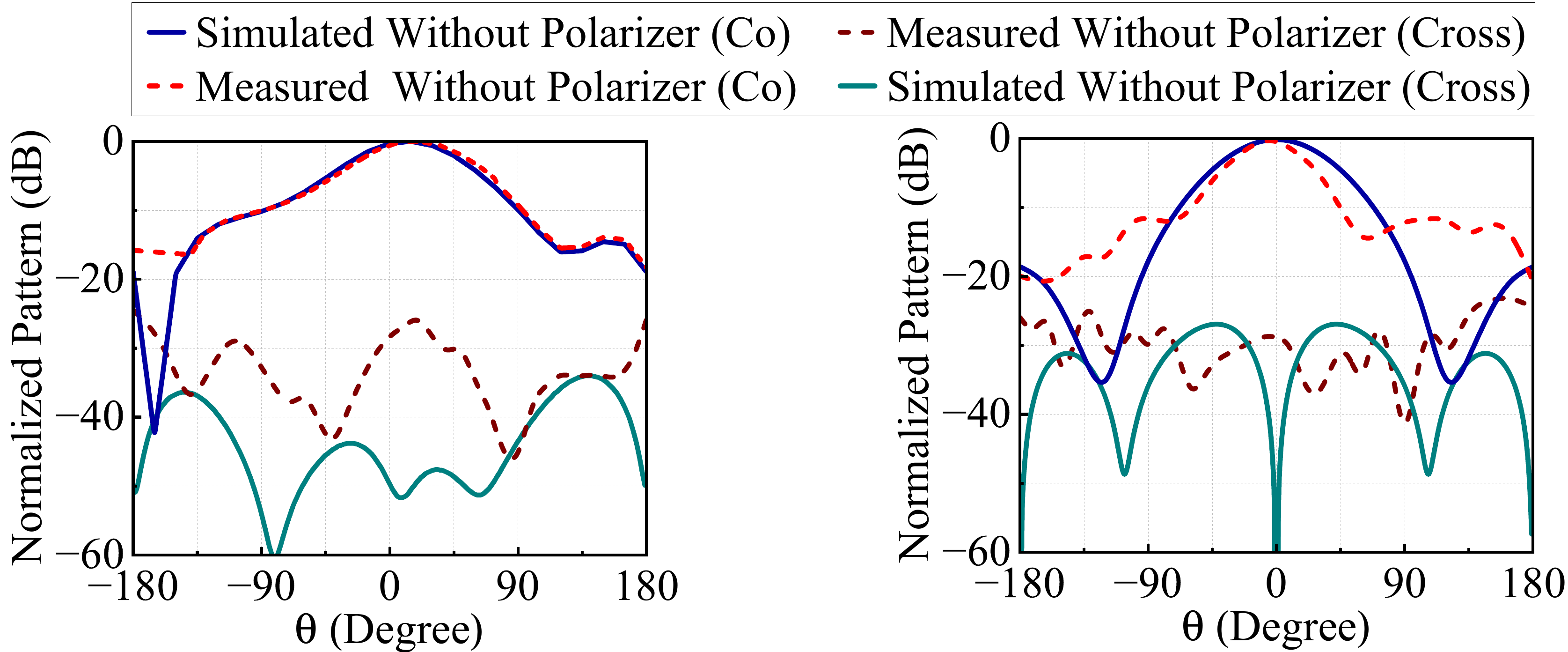}
\par\vspace{-1mm}

\makebox[\columnwidth][c]{%
    \makebox[0.5\columnwidth][c]{\footnotesize (a)}%
    \makebox[0.5\columnwidth][c]{\footnotesize (b)}%
}
\par\vspace{-1mm}

\caption{Simulated and measured far-field radiation patterns of the linear
feed source at 1 GHz in (a) $E$-plane and (b) $H$-plane.}
\label{ffpp}
\end{figure}
%%%%%%%%%%%%%%%%%%%%%
\begin{figure}[t]
\centering
\includegraphics[width=\columnwidth]{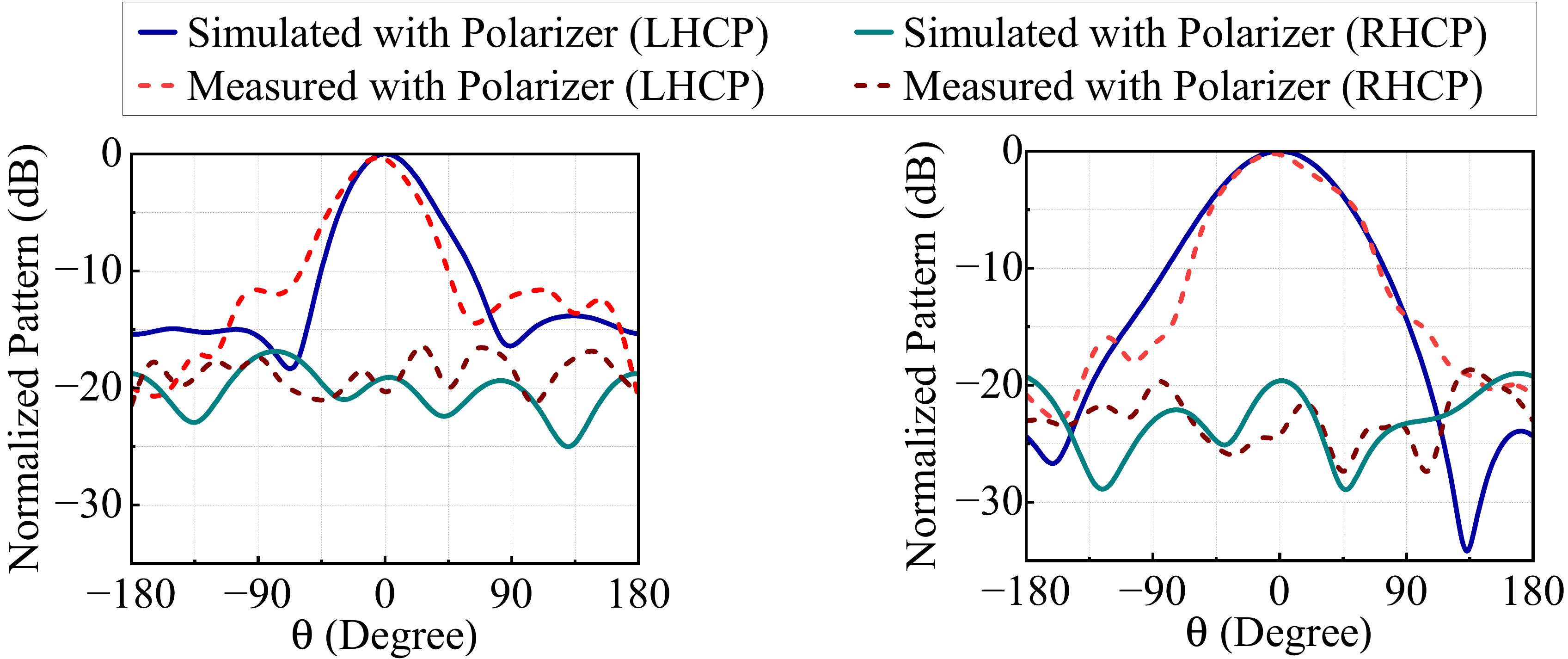}
\par\vspace{-1mm}

\makebox[\columnwidth][c]{%
    \makebox[0.5\columnwidth][c]{\footnotesize (a)}%
    \makebox[0.5\columnwidth][c]{\footnotesize (b)}%
}
\par\vspace{-1mm}

\caption{Simulated and measured far-field radiation patterns of the feed source with
polarizer at 1 GHz in (a) $xz$ plane and (b) $yz$ plane.}
\label{fig9}
\end{figure}
%%%%%%%%%%%%%%%%%%%%%%%%%%%%%%
\begin{figure}[!t]
\centering
% ---------- (a) and (b) ----------
\includegraphics[width=\columnwidth]{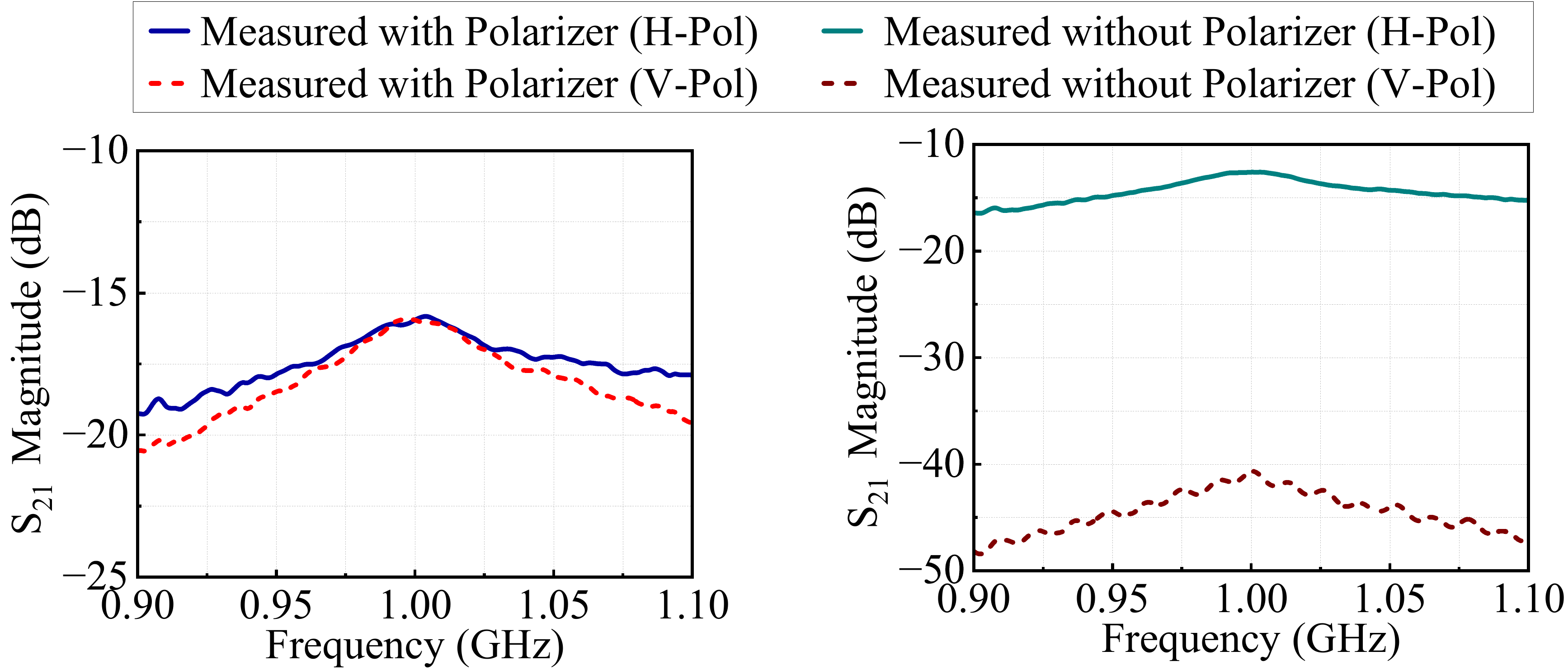}%
\vspace{-0.5mm}

\makebox[\columnwidth][c]{%
    \makebox[0.5\columnwidth][c]{\footnotesize (a)}%
    \makebox[0.5\columnwidth][c]{\footnotesize (b)}%
}%
\vspace{1.5mm}

% ---------- (c) and (d) ----------
\includegraphics[width=\columnwidth]{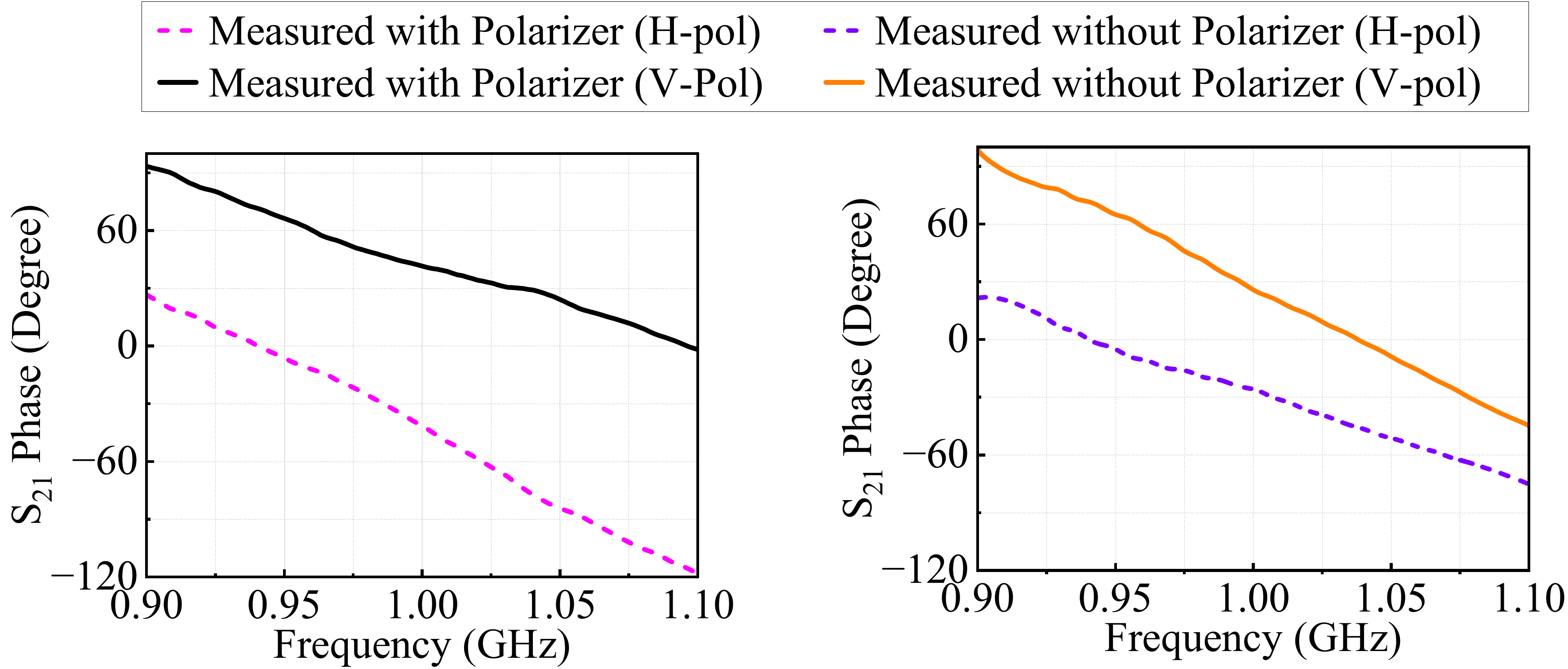}%
\vspace{-0.5mm}

\makebox[\columnwidth][c]{%
    \makebox[0.5\columnwidth][c]{\footnotesize (c)}%
    \makebox[0.5\columnwidth][c]{\footnotesize (d)}%
}%
\vspace{1.5mm}

% ---------- (e) and (f) ----------
\includegraphics[width=\columnwidth]{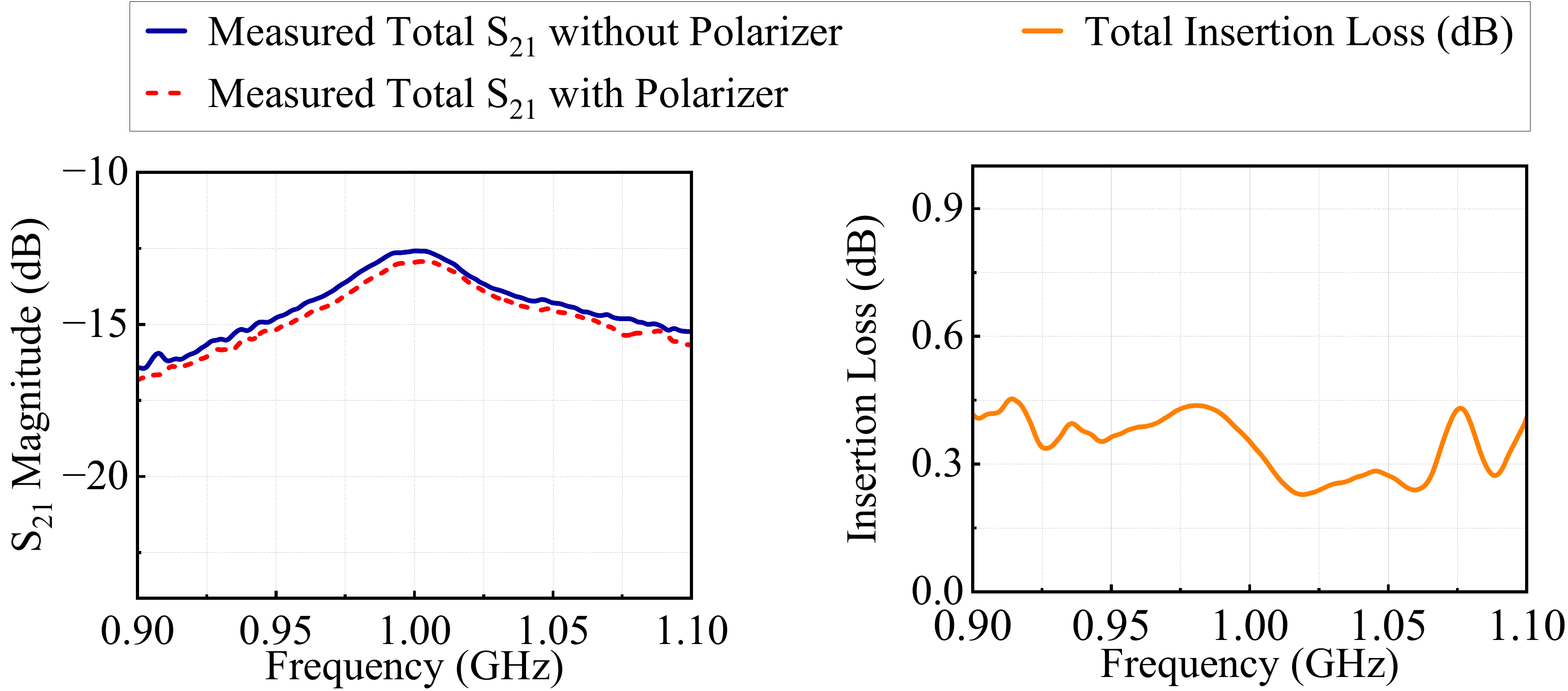}%
\vspace{-0.5mm}

\makebox[\columnwidth][c]{%
    \makebox[0.5\columnwidth][c]{\footnotesize (e)}%
    \makebox[0.5\columnwidth][c]{\footnotesize (f)}%
}%
\vspace{1mm}
\caption{Measured transmission characteristics (a) H- and V-polarized \(S_{21}\) with the polarizer, (b) H- and V-polarized \(S_{21}\) without  the polarizer, (c) transmission phases with the converter, (d) transmission phases without the polarizer, (e) total \(S_{21}\) with and without the polarizer, and (f) insertion loss.}
\label{lss}
\vspace{-.1cm}
\end{figure}
%%%%%%%%%%%%%%%
\begin{figure} [t]
\centering
\includegraphics[width=\columnwidth]{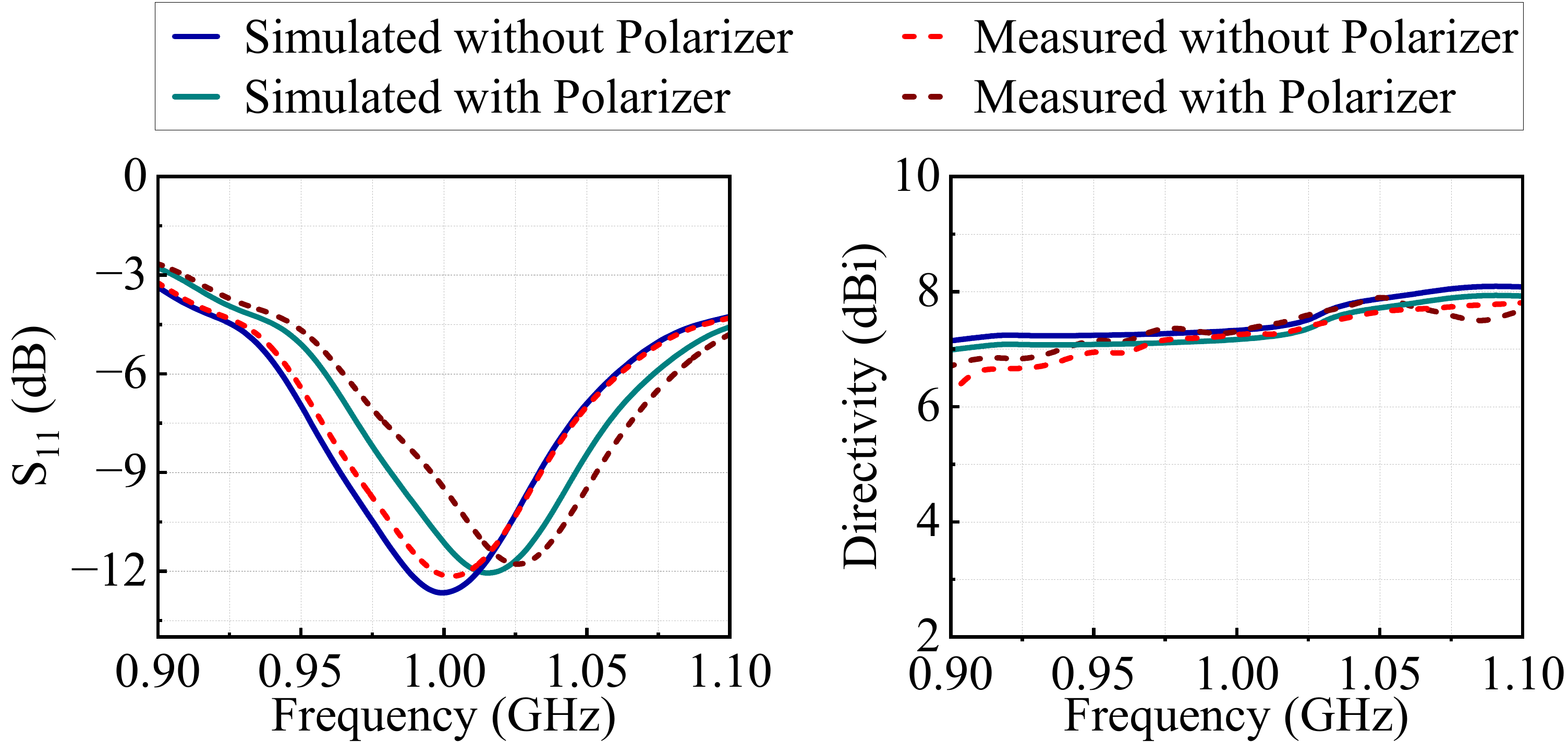}
\par\vspace{-1mm}

\makebox[\columnwidth][c]{%
    \makebox[0.5\columnwidth][c]{\footnotesize (a)}%
    \makebox[0.5\columnwidth][c]{\footnotesize (b)}%
}
\par\vspace{-1mm}

\caption{Simulated and measured results (a) reflection coefficient
and (b) directivity.}
\label{direc}
\end{figure}
%%%%%%%%%%%%%%%%
\section{Conclusion}
This work presented a miniaturized high-permittivity dielectric LP-to-CP converter for L-band operation. The in-house-fabricated dielectric medium combined a 3-D-printed PLA enclosure with a TiO$_2$–water mixture and exhibited an effective relative permittivity of 11.3 at 1 GHz. The polarizer occupied $0.83\lambda_0 \times 0.77\lambda_0 \times 0.22\lambda_0$.  The prototype exhibited a measured 3-dB axial-ratio bandwidth of 12\%, with a low insertion loss below 0.6 dB. At 1 GHz, the measured radiation patterns exhibited broadside LHCP radiation with LHCP-to-RHCP discrimination levels of 19 and 24 dB in the $xz$- and $yz$-planes, respectively. The proposed approach combines a low-profile design, permittivity adjustment, low insertion loss and low-cost additive manufacturing that enabled compact L-band air and space communication systems.
\begin{table*}[!t]
\caption{Comparison of the Proposed Converter With Recent
LP-to-CP Converters}
\label{tab:comparison}
\centering
\scriptsize
\setlength{\tabcolsep}{2.0pt}
\renewcommand{\arraystretch}{1.7}

\begin{tabular}{
>{\centering\arraybackslash}m{1.2cm}
>{\centering\arraybackslash}m{1.55cm}
>{\centering\arraybackslash}m{1.80cm}
>{\centering\arraybackslash}m{1.55cm}
>{\centering\arraybackslash}m{1.65cm}
>{\centering\arraybackslash}m{6.90cm}
>{\centering\arraybackslash}m{1.40cm}}
\hline

\textbf{Ref.} &
\makecell[c]{\textbf{Frequency}\\\textbf{(GHz)}} &
\makecell[c]{\textbf{3-dB}\\\textbf{ARBW (\%)}} &
\makecell[c]{\textbf{Insertion Loss}\\\textbf{(dB)}} &
\makecell[c]{\textbf{Polarizer}\\\textbf{Height$^{*}$}} &
\textbf{Structure} &
\textbf{Complexity} \\
\hline

\cite{geng2025broadband} &
24.45 &
52.8 &
$1.0\,/\,1.65$ &
0.55 &
Multilayer frequency-selective surface &
High \\
\hline

\cite{ouslimani2026kadual} &
20.0 / 30.0 &
10.0 / 6.7 &
$1.0\,/\,1.0$ &
0.56 / 0.84 &
Multilayer metasurface with dielectric impedance transformers &
High \\
\hline

\cite{zhao2021conformal} &
18.0 &
41.3 &
2.28 &
1.80 &
3-D-printed all-dielectric structure &
Low \\
\hline

\cite{ding2021lowprofile} &
24.0 &
25.0 &
1.2 &
0.52 &
3-D-printed dielectric structure with anisotropic
antireflection layers &
Moderate \\
\hline

\cite{Arnieri2021} &
20.0 &
24.0 &
1.1 &
0.27 &
Dual-layer Jerusalem-cross FSS &
Moderate \\
\hline

\cite{Dicandia2022} &
3.5 &
20.7 &
1.79 &
0.23 &
Two-layer fully metallic FSS &
Moderate \\
\hline

\textbf{This work} &
\textbf{1.0} &
\textbf{12.0} &
\textbf{0.6} &
\textbf{0.22} &
\textbf{Additively manufactured high-permittivity
dielectric polarizer} &
\textbf{Low} \\
\hline

\end{tabular}

\vspace{1mm}
\begin{minipage}{0.98\textwidth}
\footnotesize
$^{*}$The polarizer height is normalized with respect to the
free-space wavelength $\lambda_0$ at the center frequency.
\end{minipage}
\end{table*}
\bibliographystyle{IEEEtran}
\bibliography{references}

\end{document}